\documentclass[journal=jpc]{achemso}

\usepackage{titlesec}
\usepackage{indentfirst}
\usepackage{hyperref}
\usepackage{siunitx}
\usepackage{amsmath}
\usepackage[normalem]{ulem}
\usepackage{soul}

\title{First-principles dissociation pathways of BCl$_3$ on the Si(100)-2$\times$1 surface}

\author{Quinn T. Campbell}
\affiliation{Center for Computing Research, Sandia National Laboratories, Albuquerque, NM, USA}
\email{qcampbe@sandia.gov}
\author{Shashank Misra}
\affiliation{Sandia National Laboratories, Albuquerque, NM, USA}
\author{Jeffrey A. Ivie}
\affiliation{Sandia National Laboratories, Albuquerque, NM, USA}
\date{\today}

\begin{document}


\abstract{
One of the most promising acceptor precursors for atomic-precision $\delta$-doping of silicon is BCl$_3$. 
The chemical pathway, and the resulting kinetics, through which BCl$_3$ adsorbs and dissociates on silicon, however, has only been partially explained. 
In this work, we use density functional theory to expand the dissociation reactions of BCl$_3$ to include reactions that take place across multiple silicon dimer rows, and reactions which end in a bare B atom either at the surface, substituted for a surface silicon, or in a subsurface position. 
We further simulate resulting scanning tunneling microscopy images for each of these BCl$_x$ dissociation fragments, demonstrating that they often display distinct features that may allow for relatively confident experimental identification. 
Finally, we input the full dissociation pathway for BCl$_3$ into a kinetic Monte Carlo model, which simulates realistic reaction pathways as a function of environmental conditions such as pressure and temperature of dosing. 
We find that BCl$_2$ is broadly dominant at low temperatures, while high temperatures and ample space on the silicon surface for dissociation encourage the formation of bridging BCl fragments and B substitutions on the surface. 
This work provides the chemical mechanisms for understanding atomic-precision doping of Si with B, enabling a number of relevant quantum applications such as bipolar nanoelectronics, acceptor-based qubits, and superconducting Si.
}

\maketitle

\section{Introduction}\label{sec:intro}
Atomic precision advanced manufacturing (APAM) techniques can be used to realize a wide variety of novel quantum and electronic devices in silicon \cite{ward2020atomic,bussmann2021atomic,schofield2025roadmap}. 
In this technique, a hydrogen or halogen terminated silicon surface is exposed to a precision lithography tool (usually, but not always,\cite{katzenmeyer2021photothermal,constantinou2024euv} scanning tunneling microscopy (STM)), which depassivates a region of interest in ultrahigh vacuum (UHV).
This region is then exposed to a precursor gas, which selectively adsorbs onto the exposed silicon, but not the surrounding resist. 
The precursor gas will dissociate on the surface, ideally leading to incorporation of the dopant atom, often at carrier densities exceeding \SI{1e14}{cm^{-2}}\cite{keizer2015suppressing}.
Most of the work on APAM so far has focused on using a precursor gas of phosphine for donor doping of silicon\cite{schofield2003atomically,ruess2007realization,fuechsle2012single,wang2022experimental}, but there has been recent interest in expanding the range of applications available by introducing alternative donors such as As\cite{stock2020atomic,stock2024single}, as well as a variety of precursors with the goal of acceptor incorporation such as diborane (B$_2$H$_6$) \cite{vskerevn2020bipolar,campbell2021model}, aluminum tricholoride (AlCl$_3$) \cite{radue2021alcl3}, boron trichloride (BCl$_3$) \cite{silva2020reaction,dwyer2021b}, and organics such as trimethyl and triethyl aluminum \cite{owen2021alkyls}. 

Of these, BCl$_3$ has shown particular promise as an APAM acceptor precursor, achieving an active carrier concentration of \SI{1.9e14}{cm^{-2}} \cite{dwyer2021b}.
To make further progress on creating quantum devices with this precursor, it is necessary to develop a comprehensive understanding of how BCl$_3$ decomposes on the surface. 
Campbell \textit{et al.} used density functional theory (DFT) to establish the initial decomposition pathway of BCl$_3$ on the silicon surface,\cite{campbell2022reaction} demonstrating that the BCl$_3$ molecule adsorbs without a reaction barrier, and quickly decomposes with low reaction barriers ($\sim$ 1 eV).
Based on this analysis, it was then predicted that BCl$_3$ doping under certain conditions could be used to achieve deterministic single acceptor doping with atomic precision $\pm$ 1 lattice site \cite{campbell2022hole}. 

This analysis was limited, however, by the aim of single acceptor incorporation. 
We only considered BCl$_x$ decomposition pathways taking place entirely along a single dimer row and ending with BCl fragments.
This paper extends these reactions to take place across multiple silicon dimers and reach an isolated B, providing the theoretical basis for predicting BCl$_x$ decomposition pathways which end in multiple acceptors incorporated in the silicon, particularly $\delta$-doping. 
This will be useful for predicting ideal APAM conditions for maximizing B incorporation, with useful applications such as improved coherence times in multi-atom qubits\cite{hsueh2014spin,watson2017atomically} and superconducting Si regions\cite{shim2014bottom}.
Furthermore, comparing these dissociation pathways with experimentally observed BCl$_3$ dissociation on the silicon surface is difficult without generation of STM images matching configurations.
Wyrick \textit{et al.} have recently generated STM images for the full range of PH$_x$ fragments on the Si(100)-2$\times$1 surface and have achieved good agreement with experimental characterization \cite{wyrick2022enhanced}.
This work will enable similar experimental analysis of BCl$_x$ fragments on Si.

In this paper, we use DFT to further expand the possible range of BCl$_x$ decomposition reactions, including reactions that take place across multiple silicon dimer rows and reactions that lead to subsurface B. 
We simulate STM images corresponding to common BCl$_x$ fragments on the surface, which can be compared with experimental results for identification purposes. 
We then develop a kinetic Monte Carlo (KMC) simulation, which allows us to predict the most common BCl$_x$ fragments during decomposition as a function of BCl$_3$ dose pressure, time, and temperature, as well as the following anneal temperature. 
We predict surface coverages of B roughly in line with experimentally observed acceptor densities. 
We further predict that BCl$_2$ will be the dominant surface feature during typical $\delta$-doping conditions.
This is largely limited by the space to decompose, however, and in systems where there is more room to dissociate, bridging BCl is often observed, with the most common location for a single B to end as a substitution for a surface silicon. 
With this knowledge, atomic precision doping of silicon using BCl$_3$ can be fine tuned to provide optimal electronic properties for desired applications.

\section{Methodology}\label{sec:methods}
\subsection{Reaction Barrier Calculations}

To compute reaction barriers between configurations, we use the Nudged Elastic Band (NEB) method, as implemented in {\sc quantum-espresso}\cite{Giannozzi2009}, and use norm-conserving pseudopotentials from the PseudoDojo repository\cite{VanSetten2018} and the Perdew-Burke-Ernzerhof exchange correlation functional.\cite{Perdew1996}
Kinetic energy cutoffs of 50~Ry and 200~Ry are used for the plane wave basis sets used to describe the Kohn-Sham orbitals and charge density, respectively, and a 2$\times$2$\times$1 Monkhorst-Pack grid is used to sample the Brillioun zone.\cite{monkhorst1976special}

BCl$_x$ fragment calculations are performed on a 4$\times$4 supercell of a seven-layer thick Si(100)--2$\times$1 slab with a 20~\AA~vacuum region, where a single unit cell has a size of 3.87~\AA $\times$ 3.87~\AA.
Matching the experimentally measured Si structure, we model the Si surface with alternating buckled Si dimers.
On the other end of the slab, the Si dangling bonds are passivated with Se atoms to prevent spurious surface effects.
The bottom four layers of the slab are frozen and the geometry of the surface layers along with the adsorbate are relaxed until the interatomic forces are lower than 50~meV/\AA.

\subsection{Scanning Tunneling Microscopy Calculations}
Simulated Scanning Tunneling Microscopy (STM) images of the configurations calculated were generated by {\sc quantum-espresso}'s postprocessing methodologies, which use the Tersoff-Hamann approximation \cite{tersoff1985theory}. 
We simulate constant current STM images, with a charge threshold of \SI{5e-5}{e} and the voltages stated in our figures.
It should be noted that these voltages are likely to somewhat underestimate real world energies \cite{stowasser1999kohn}, and that this approximation does not take into account the effect of band bending on the substrate from the electric field introduced by the STM tip.
In our calculations of reaction barriers for the dissociation of BCl$_x$, the silicon dimers on our surface have an alternating tilt up and down-- the lowest energy reconstruction.
When we simulate STM images, however, we hold all the silicon dimers which are not directly connected to the relevant fragment flat.
This mimics the rapid oscillation between tilt directions which leads to silicon dimers looking flat in experimentally obtained STM images \cite{engelund2016butterfly}.
In the main paper, we simulate STM images at normal resolution, but in the Supporting Information with all the simulated STM images, we also include images with an added gaussian blur to mimic realistic experimental noise. 
This approach mimics the STM simulation techniques used by Wyrick \textit{et al.} in their analysis of PH$_x$ fragments on the silicon surface \cite{wyrick2022enhanced}.

\subsection{Kinetic Monte Carlo Simulations}

Finally, we use a kinetic Monte Carlo (KMC) model \cite{Bortz1975KMC,Gillespie1976KMC} as implemented in the KMCLib package \cite{Leetmaa2014KMClib} to determine the probability of incorporation.
Our KMC model uses transition rates based on the Arrhenius equation $ \Gamma = A \exp{\Delta/k_{\rm B}T}$~\cite{arrhenius1889reaktionsgeschwindigkeit}, where $\Gamma$ is transition rate, $A$ is the attempt frequency, $\Delta$ is the reaction barrier found from our earlier DFT calculations, $k_{\rm B}$ is the Boltzmann constant, and $T$ is the temperature. 
We set all attempt frequencies $A$ to $10^{12}$ s$^{-1}$ as a reasonable order of magnitude estimate based on an analysis of attempt frequencies for the dissociation of phosphine on silicon~\cite{warschkow2016reaction}, which we assume to be analogous. 
We calculate the effusive flow rate of molecules landing on any particular silicon dimer as $\Phi_{effusion} =PA/\sqrt{2\pi m k_{\rm B}T}$, where $P$ is the pressure of the incoming precursor gas, $A$ is the area of impingement, taken here as a single silicon dimer, $m$ is the mass of the precursor gas.

Each KMC calculation is repeated 100 times with different random seeds, and the sample mean of the results is reported. 
The code for these simulations has been uploaded to an open source repository\cite{kmccode}.

\section{Results}\label{sec:results}
\subsection{Reaction Barriers}
\label{subsec:rxn-barrs}
We begin by calculating novel reaction barriers for BCl$_x$ dissociation on the silicon surface.
These reactions expand on our previous work \cite{campbell2022reaction} in three main directions. 
First, we consider reactions which take place across multiple silicon dimer rows. 
This provides another avenue for Cl to shed from the initial BCl$_3$ molecule and sometimes even for BCl$_x$ fragments to switch dimer rows as shown in Fig.~\ref{fig:rxns1}.
Second, we discover a new configuration of BCl which has not been previously identified, a bridging BCl which is located entirely along one silicon dimer (see, e.g., Fig.~\ref{fig:rxns1}e). 
This is analogous to the previously seen location of PH molecules \cite{warschkow2016reaction}.

\begin{figure*}
\includegraphics[width=0.9\textwidth]{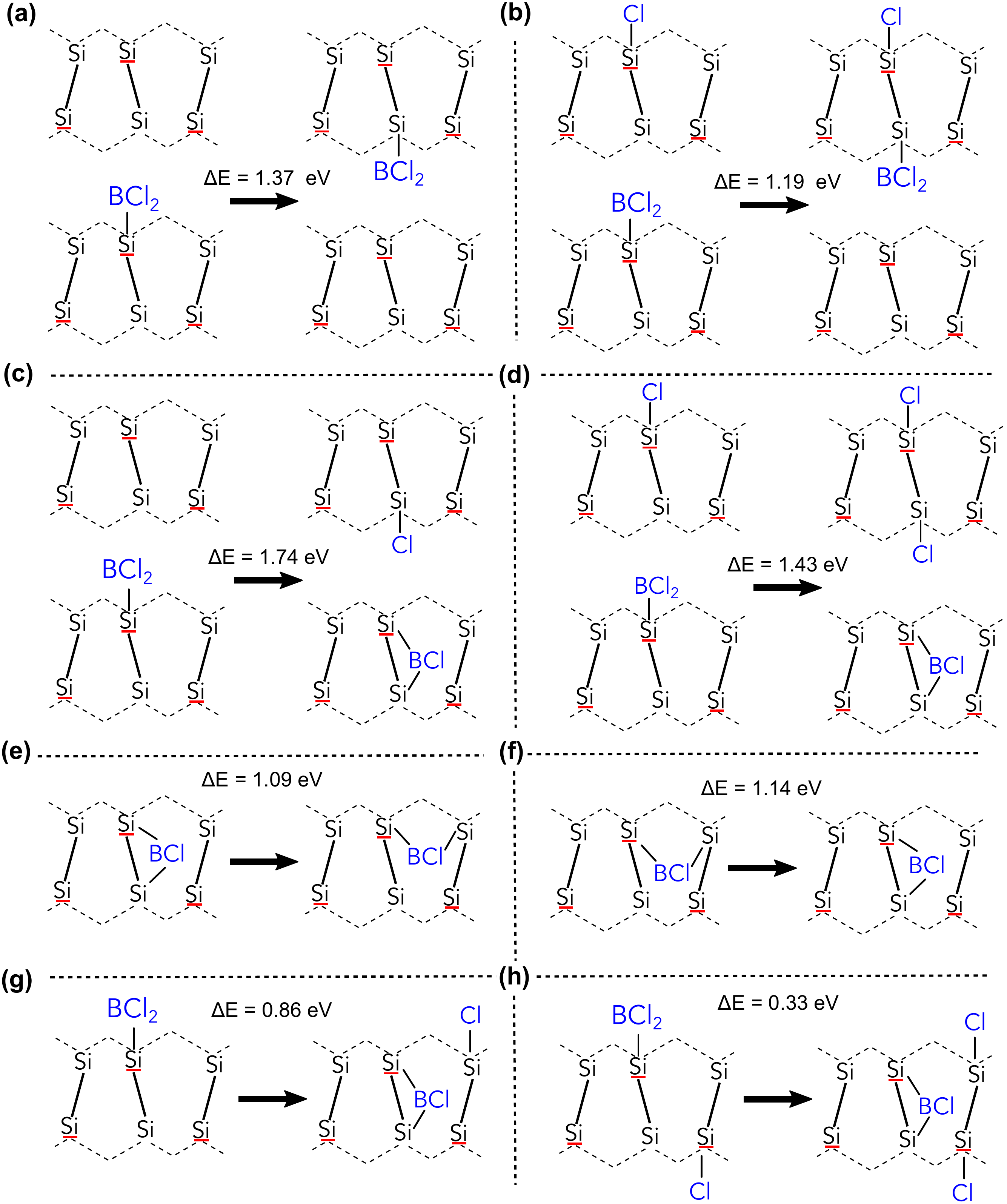}

\caption{Schematic diagrams of reaction barriers for BCl$_x$ fragments dissociating on the Si surface. Panels a-d depict reactions which take place across multiple silicon dimer rows, while e-h represent single silicon dimer row reactions. Silicon atoms which have red underlines represent the lower end of the tilted silicon dimer. $\Delta$E represents the reaction barrier of each reaction. \label{fig:rxns1}}
\end{figure*}

Finally, we extend our analysis beyond BCl fragments to look at decomposition into individual B atoms in Fig.~\ref{fig:rxns3}.
The range of possible locations for a single B  can be quite high as it migrates from residing solely on the surface to subsurface locations.
We chose sites for B incorporation matching previous results for the initial incoporation of P into the Si surface \cite{warschkow2016reaction}.
While the locations we tested are not comprehensive, we looked at B locations both adsorbed on the surface often in similar configurations as previously shown BCl$_x$ fragments (as in Fig.~\ref{fig:rxns3}a and b), one layer within the subsurface of the material (Fig.~\ref{fig:rxns3}c and d), and B substituting for a silicon atom on the surface layer (Fig.~\ref{fig:rxns3}e and f).
Reaction barriers for reaching an isolated B are of the order $\approx$ 1.5 eV, well within the range of typical reaction barriers seen in the dissociation steps up to BCl. 
We therefore expect that BCl$_3$ dissociation all the way to individual B atoms on the surface will be possible at reasonable anneal temperatures (e.g. \SI{350}{\celsius}).

\begin{figure*}
\includegraphics[width=0.9\textwidth]{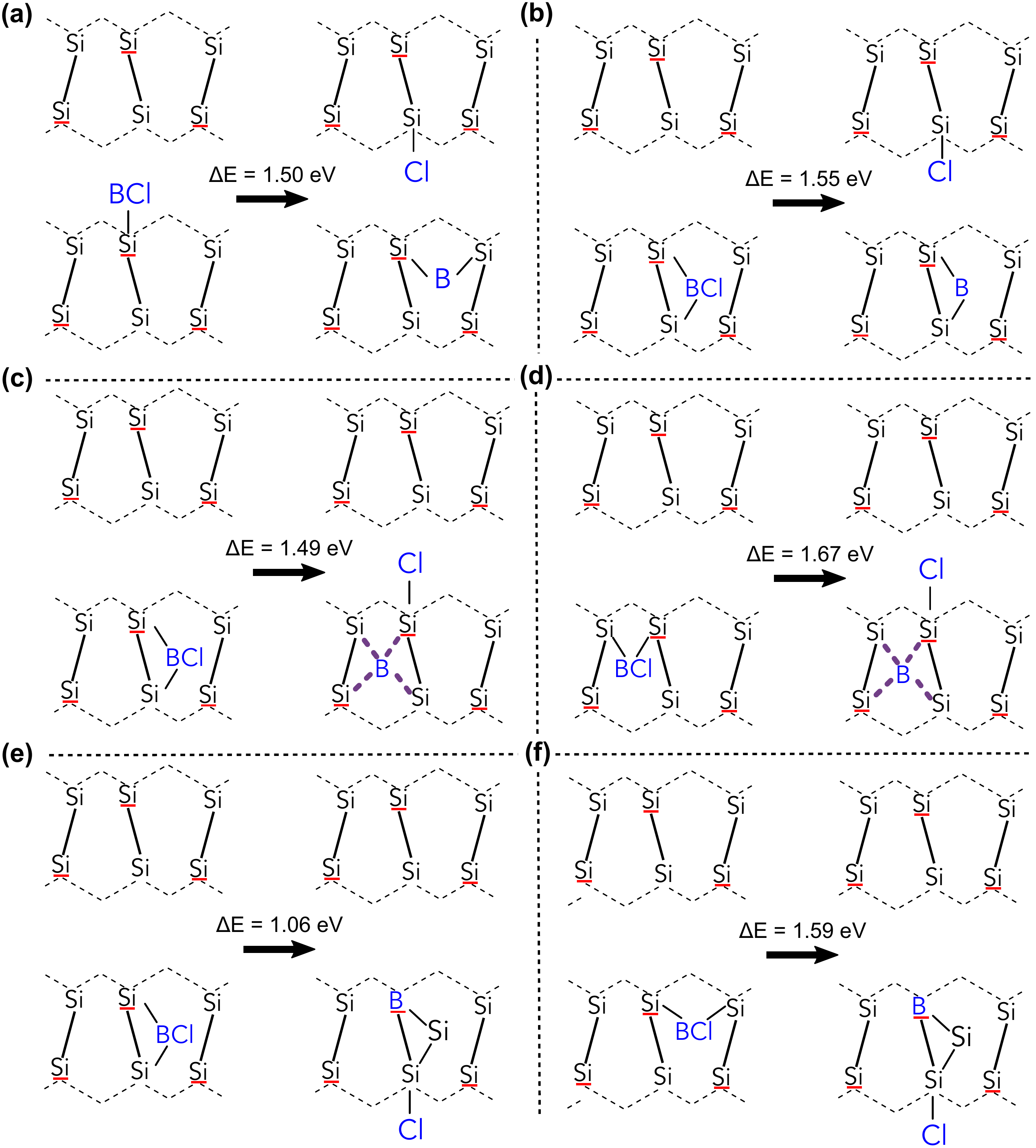}

\caption{Continued schematic diagrams of reaction barriers for BCl$_x$ fragments dissociating on the Si surface. Silicon atoms which have red underlines represent the lower end of the tilted silicon dimer. The dashed purple lines connecting to the B atom in panels c and d indicate that the B is actually subsurface in these models. See Fig.~\ref{fig:stm} for a ball and stick diagram of the B location. \label{fig:rxns3}}
\end{figure*}

Due to the combinatorial explosion of possible reactions once the boron reaches the subsurface, we have stopped our modeling of reaction barriers at this first level of boron atoms at or slightly below the surface.
These systems all exhibit extremely strong ($<$ -3.0 eV) adsorption energies and high reverse reaction barriers (i.e. the barrier of moving the reaction in the opposite direction as shown within Figs.~\ref{fig:rxns1}-~\ref{fig:rxns3}).
We therefore assume within our models that once a BCl$_3$ molecule dissociates to the point of a single B atom, that B atom will remain on the surface throughout processing and for subsequent electrical measurements, will behave as an electrically active dopant.

Additional reaction barriers can also be found in the supporting information.

\subsection{STM simulations of surface configurations}
\label{subsec:stm-sim}
\begin{figure*}
\includegraphics[width=\textwidth]{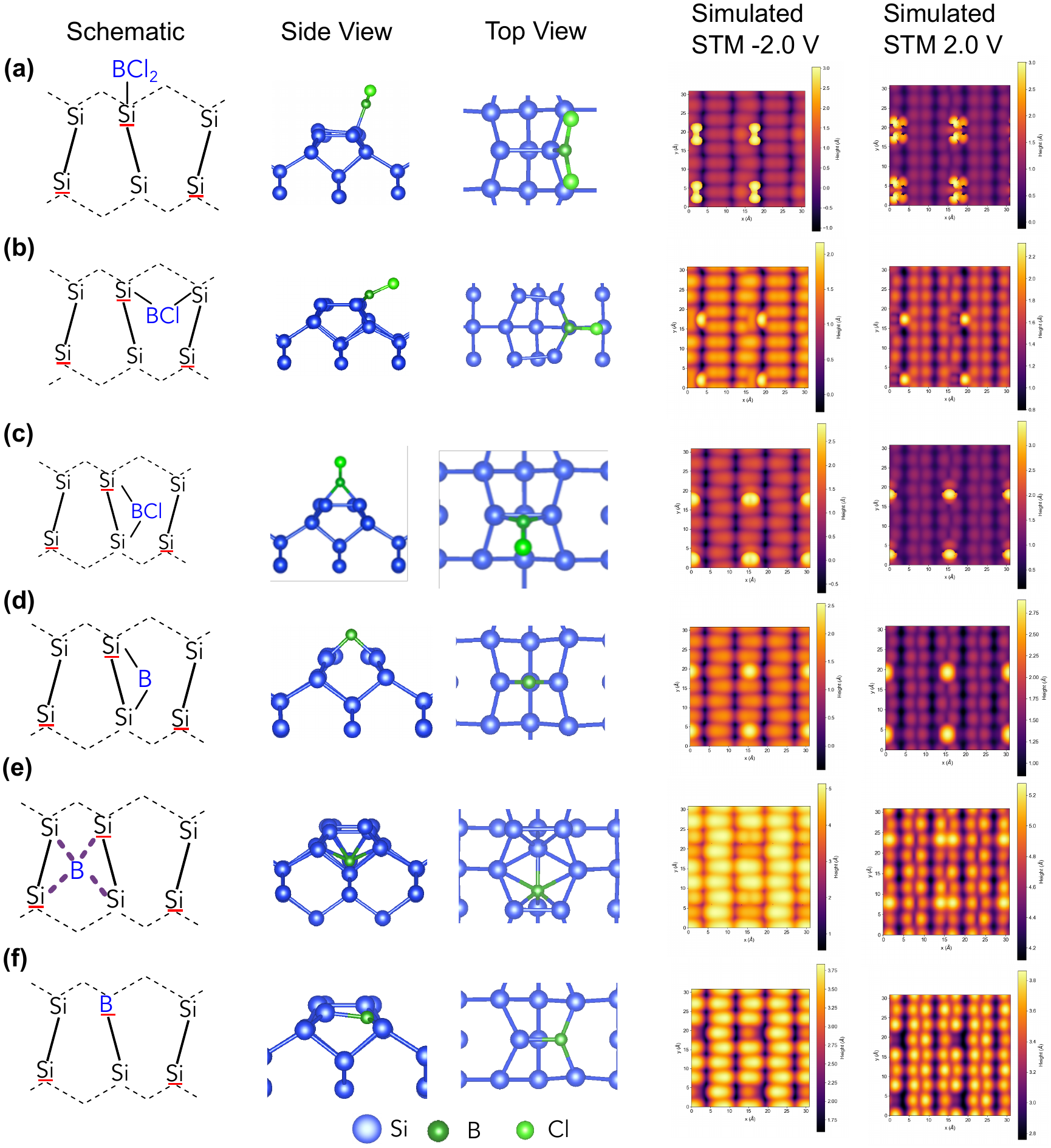}
\caption{ Scanning tunneling microscopy (STM) simulations of BCl and B fragments that serve as likely endpoints for BCl$_x$ dissociation.
In each row, we include the schematic view of what the BCl$_x$ fragment looks like in the style of Figs.~\ref{fig:rxns1}-\ref{fig:rxns3}, a ball and stick view of the fragment from the top and the bottom, and finally the simulated STM image at -2.0 V and +2.0 V.
\label{fig:stm}}
\end{figure*}

To aid in the experimental identification of the predicted BCl$_x$ configurations, we have performed STM simulations of all BCl$_x$ configurations.
In Fig.~\ref{fig:stm}, we display the STM simulations for the most likely end points of BCl$_x$ decomposition, i.e. bridging BCl and B fragments, as well as subsurface B, and B substitutions into the Si lattice.

Fortunately for later experimental verification, the majority of the simulated STM structures appear to have distinct features which could be used for identification. 
BCl$_2$, shown in Fig.~\ref{fig:stm}a, has a bright barbell like pattern.
To contrast, the bridging BCl fragment across multiple dimer rows, shown in Fig.~\ref{fig:stm}b, has barbell like patterns where a bright Cl atoms is immediately next to a darker depression (this pattern is particularly pronounced in the positive bias images).
The STM patterns with the most similarity are the bridging BCl on a single Si dimer (Fig.~\ref{fig:stm}c) and the bridging B along a single Si dimer (Fig.~\ref{fig:stm}d)
Both display a single bright spot corresponding to the elevated atoms.
In the bridging BCl case, the Cl atom is slightly offset from the silicon dimers, however, and the bright spot is thus centered between the silicon dimer rows.
In contrast, the bridging B is entirely centered on the silicon dimer and the bright spot aligns neatly with the nearby dimers. 
This means that it should be possible to distinguish between BCl$_2$, BCl and a bare B, even if they are occupying essentially the same position along the silicon surface. 
The subsurface B, in Fig.~\ref{fig:stm}e is the hardest to distinguish in our simulations, with the majority of the surface appearing to mimic a normal bare silicon surface. 
The subsurface B can be identified in negative bias images solely by a slightly less bright silicon dimer, whereas at positive bias, we begin to see a small gap between the relevant top silicon atoms.
Finally, for a B substitution, shown in Fig.~\ref{fig:stm}f for a Si atom on the surface, the B atom appears significantly less bright than the surrounding atoms, which is again clearer at positive bias.
When switching from negative to positive bias, the observed trends are only exaggerated at lower and higher biases, as can be seen in the supporting information.
These results emphasize the utility of taking multiple voltages for STM images when attempting to identify specific BCl$_x$ fragments on the surface. 

In the supporting files for this work, we include STM simulation data and images for all the 26 BCl$_x$ dissociation fragments we have simulated.

\subsection{Kinetic Monte Carlo Simulations of BCl$_x$ }
\label{subsec:kmc_results}

Finally, we take the reaction barriers calculated from Section \ref{subsec:rxn-barrs} and input them (along with the previously calculated reaction barriers for BCl$_3$\cite{campbell2022reaction}) into a kinetic Monte Carlo (KMC) simulation. 
This method works by surveying the available reactions from an initial setup, enumerating all possibilities from the reaction list, and then choosing a reaction from the probability distribution determined by the reaction barriers (as well as a given temperature the simulation takes place at).
This allows us to predict how BCl$_3$ will dissociate on the surface of silicon accounting for real stochasticity of kinetic incorporation \cite{ivie2021impact}.
Our aim with these simulations is to predict what will be the dominant fragments of BCl$_3$ on the silicon surface after a realistic dosing and incorporation schedule. 

\begin{figure*}
    \centering
    \includegraphics[width=\linewidth]{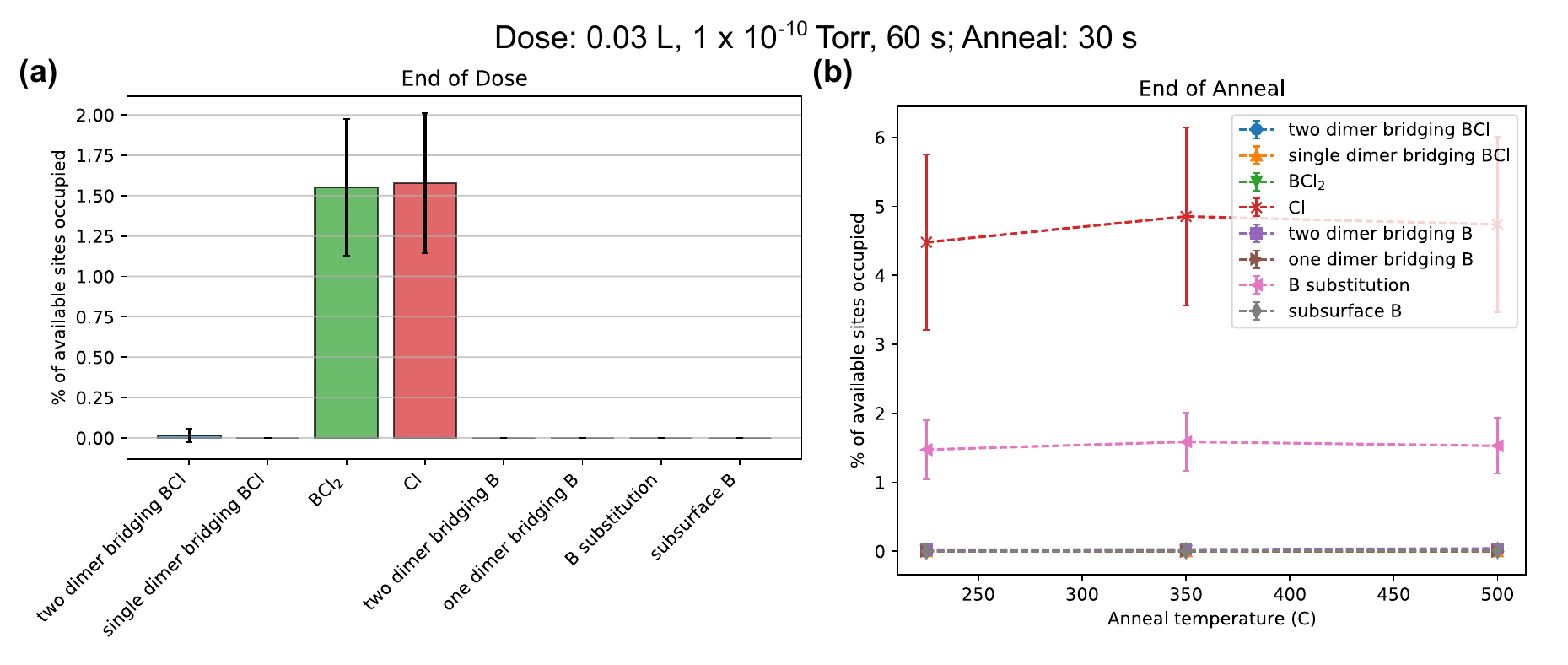}
    \caption{The percentage of the silicon surface taken up by different BCl$_x$ fragments at the end of the (a) dose and (b) anneal, with a dependence on the anneal temperature. These results come from an initial dosing of 0.03 L, with a dosing pressure of  \SI{5e-10}{Torr} for 60 s at room temperature followed by a 30 s anneal at varying temperatures.}
    \label{fig:0.03L-kmc}
\end{figure*}

We initially simulate a system at low coverage, 0.03 L, using a dosing pressure of \SI{5e-10}{Torr} for 60 s at room temperature, followed by an anneal at temperatures of \SI{225}{\celsius}, \SI{350}{\celsius}, or \SI{500}{\celsius} for 30 s.
This allows us to examine what dissociation fragments we expect to see when each of the BCl$_3$ molecules is allowed to fully dissociate without necessarily running into steric hindrance from nearby BCl$_x$ fragments.
In Fig.~\ref{fig:0.03L-kmc}, we show that bridging BCl$_2$ and Cl fragments tend to dominate on the surface at the end of the dosing run at nearly equal proportions.
This can be attributed to the fact that the first Cl atom automatically detaches from the BCl$_2$ in the process of adsorbing onto the silicon substrate \cite{campbell2022reaction}, thus creating an equal number of BCl$_2$ and Cl molecules. 
Beyond this point, however, higher temperatures are needed to further dissociate the molecule. 
After annealing, the BCl$_2$ molecules fully dissociate into individual B atoms, which are most likely to take the form of a substitution into the silicon lattice. 
By comparing between the dose and anneal plots, it becomes clear that essentially all of the BCl$_2$ on the surface is being converted to B substitutions.
This tendency toward B substitution can be attributed to the comparatively low barrier (1.06 eV) for a substitution, compared to all other single B reactions, which require overcoming barriers $>\approx$ 1.5 eV.
Furthermore the 2 Cl atoms of the BCl$_2$ fragment are shed and the percentage of Cl on the surface is approximately three times the amount of B.
This matches the initial stoichiometric ratio of BCl$_3$ and indicates that Cl$_2$ desorption is not yet reached at these anneal temperatures within our model. 

\begin{figure*}
    \centering
    \includegraphics[width=\linewidth]{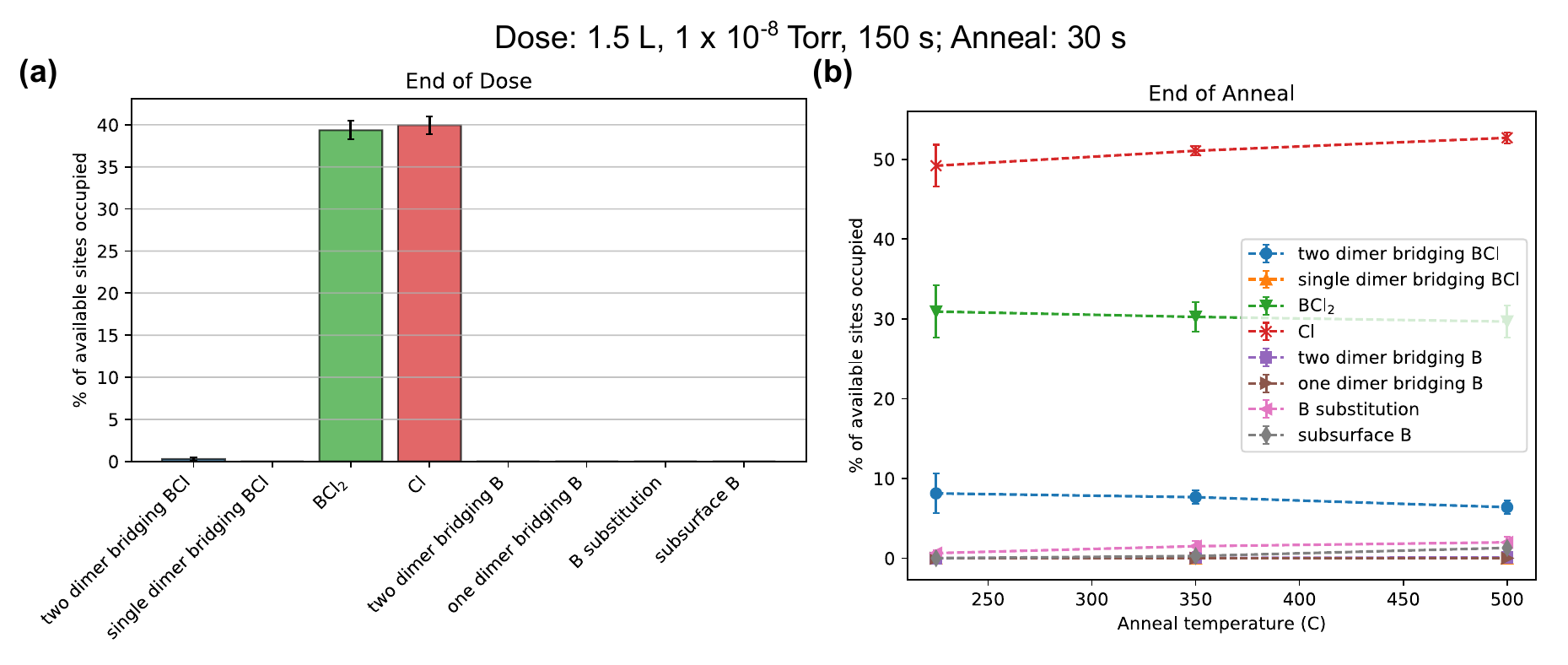}
    \caption{The percentage of the silicon surface taken up by different BCl$_x$ fragments at the end of the (a) dose and (b) anneal, with a dependence on the anneal temperature. These results come from an initial dosing of 1.5 L, with a dosing pressure of  \SI{1e-8}{Torr} for 150 s at room temperature followed by a 30 s anneal at varying temperatures.}
    \label{fig:1.5L-kmc}
\end{figure*}

\begin{figure*}
    \centering
    \includegraphics[width=\linewidth]{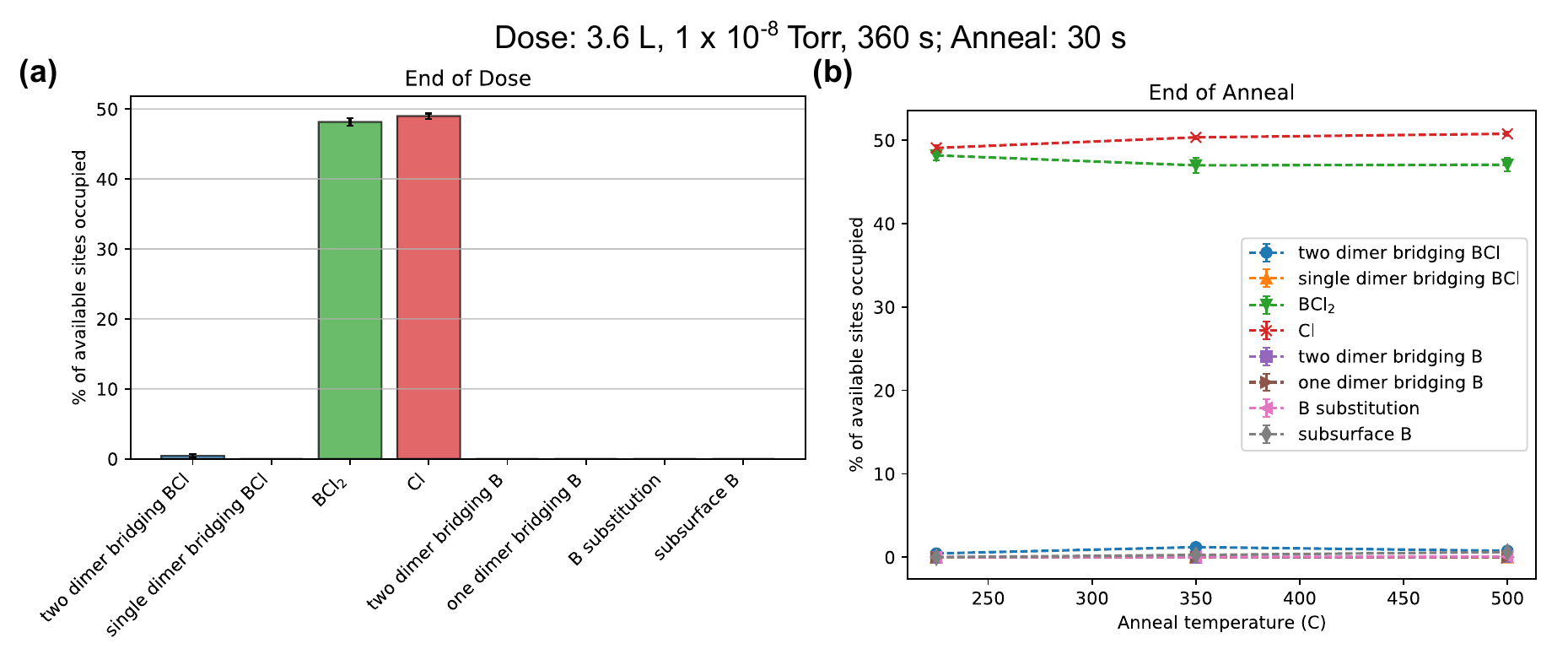}
    \caption{The percentage of the silicon surface taken up by different BCl$_x$ fragments at the end of the (a) dose and (b) anneal, with a dependence on the anneal temperature. These results come from an initial dosing of 3.6 L, with a dosing pressure of \SI{1e-8}{Torr} for 360 s at room temperature followed by a 30 s anneal at varying temperatures.}
    \label{fig:3.6L-kmc}
\end{figure*}

We then move on to simulating the formation of a $\delta$-doped layer of B, using realistic monolayer coverages of BCl$_3$. 
We follow the settings of Dwyer \textit{et al.} \cite{dwyer2021b}, using a dosing pressure of \SI{1e-8}{Torr} with different times to create a 1.5 L coverage in Fig.~\ref{fig:1.5L-kmc} and 3.6 L coverage in Fig.~\ref{fig:3.6L-kmc}. 

The most notable difference in these systems is that steric hindrance becomes a significant factor in the endpoint of BCl$_3$ dissociation.
At coverages $>$ 1 L, the number of sites that a BCl$_2$ molecule can dissociate into decreases rapidly.
As a BCl$_3$ adsorbs onto the silicon surface, it will automatically shed a Cl atom to the nearby silicon atom on the same dimer, forming BCl$_2$. 
Beyond this initial adsorption, however, it can be difficult to find room to shed the remaining Cl when most of the nearby sites are also occupied by similar BCl$_2$ + Cl pairs.
We do include Cl desorption in our model as an option, but with a reaction barrier of 4.85 eV, this does not become a significant factor until high temperature anneals. 
This means that each adsorption site inherently limits the number of sites that can be dissociated into by further BCl$_2$ molecules.
In Fig.~\ref{fig:1.5L-kmc}, we see that this initial adsorption process is dominant during the room temperature dosing, with the only significant BCl$_x$ fragment found after dosing being BCl$_2$.
This matches what we previously saw for 0.03 L dosing.
When the anneal is applied, however, the BCl$_2$ fragments have significantly less room to dissociate than in the 0.03 L case, due to the greater saturation of dosing sites from the higher Langmuir dose.
After the anneal, the majority of fragments are still BCl$_2$; however, the percentage of the surface occupied by BCl$_2$ moves from $\approx$30\% to $\approx$23\%. 
The BCl$_2$ that does dissociate primarily moves into the form of two dimer bridging BCl, at around 5\% of the surface.
This necessarily results in an increase in the percentage of Cl on the surface from $\approx$ 40\% to $\approx$ 50\%, filling the remaining surface sites. 
As anneal temperatures increases, however, the number of single B increases slightly, also slightly increasing the proportion of Cl.
The dominant form of single B is still B substitutions into the silicon surface, but at 500 C, there is a non-negligible amount of subsurface B.
Comparing this result to 0.03 ML coverage, we can infer that B substitution is the energetically easiest path, but as the surface sites saturate, moving to the subsurface becomes a viable option.

These same trends are amplified in the 3.6 L dosing. 
Here BCl$_2$ takes up a greater percentage of the surface, $\approx$45 \%. 
By the end of the dose, the BCl$_2$ and Cl sites take up essentially the entirety of the surface. 
This means, however, that there is essentially no room for the BCl$_2$ to dissociate, leading to the percentage of BCl$_2$ on the surface staying almost exactly the same after annealing.
This indicates that if the goal is to maximize dissociation to B, it may be preferable to perform sub-Langmuir doses, as this provides more room for eventual dissociation. 
Another avenue, explored in previous work for phosphine \cite{ivie2021impact}, may be to perform the dose at higher temperatures, which encourages dissociation reactions to take place during the dosing phase, before other precursor molecules have saturated the surface. 

Notably, both the 1.5 L and 3.6 L doses create a total coverage of different BCl$_x$ fragments at $\approx$ 0.5 ML at the end of the anneal.
These numbers are roughly double the reported sheet densities of Dwyer \textit{et al.} \cite{dwyer2021b} of \SI{1.9e14}{cm^{-2}} (0.26 ML).
Given that many of these fragments are still in the form of BCl$_2$, which has the relatively weakest adsorption energies of all BCl$_x$ fragments, it is somewhat likely that not all of them will electrically incorporate during the subsequent silicon growth.
Furthermore, B is known to dimerize in silicon, which renders it electrically inactive\cite{tarnow1992theory}. 
These mechanisms may account for the diminished values seen for electrically active B in experiment compared to our results.

While we have focused on $\delta$-layer formation for these kinetic simulations, the associated code \cite{kmccode} could also be used to generate the probability of incorporation for single or few acceptor systems, directly impactful for quantum devices. 

\section{Conclusions}\label{sec:conclusions}

In this work, we used DFT calculations to explore a significantly expanded dissociation pathway for BCl$_3$ on the Si(100)-2$\times$1 surface. 
We find significant new configurations which had not been previously explored such as the single Si dimer bridging BCl and also expand to dissociation to a lone B atom.
This dissociation pathway remains achievable with modest reaction barriers, usually on the order of 1.5 eV.
We furthermore explore what each of these BCl$_x$ fragments would look like on the surface under STM, finding distinct features for each. 
Finally, we use these reaction barriers to develop a kinetic model that allows us to predict how BCl$_3$ actually dissociates on the surface and what are the dominant end points.
After room temperature dosing, BCl$_2$ fragments are by far the most common.
After an anneal, if the BCl$_2$ has room to fully dissociate, such as the case of 0.03 L dosing, then the most likely outcome will be a B substitution into the silicon lattice.
If the BCl$_2$ becomes trapped by neighboring fragments, however, as in the 1.5 L and 3.6 L dosing cases, it is much less likely to fully dissociate after an anneal, with only a minority reaching substitutional boron. 
This is analogous to previous work with phosphine \cite{ivie2021impact}, which has shown that dosing at lower pressures for longer times and/or higher temperatures can be used to increase the rate at which phosphine dissociates before being hampered by adsorption of PH$_x$ fragments at nearby sites.
This work allows us to fully understand the dissociation pathway of BCl$_3$ on silicon and allows for advances in the creation of a number of unique quantum applications such as bipolar nanoelectronics, acceptor-based qubits, and superconducting Si.

\begin{acknowledgement}\label{sec:acknowledgement}
    We would like to thank Azadeh Farazeh, Robert Butera, Chris Allemang, Devika Mehta, Tom Sheridan, and Scott Schmucker for useful discussions surrounding the results of this work. 
    
    This work was supported by the LPS Qubit Collaboratory and by the Laboratory Directed Research and Development (LDRD) program at Sandia National Laboratories under project 229375.
    This work was performed, in part, at the Center for Integrated Nanotechnologies, an Office of Science User Facility operated for the U.S. Department of Energy (DOE) Office of Science.
    Sandia National Laboratories is a multi-mission laboratory managed and operated by National Technology \& Engineering Solutions of Sandia, LLC (NTESS), a wholly owned subsidiary of Honeywell International Inc., for the U.S. Department of Energy’s National Nuclear Security Administration (DOE/NNSA) under contract DE-NA0003525. This written work is authored by an employee of NTESS. The employee, not NTESS, owns the right, title and interest in and to the written work and is responsible for its contents. Any subjective views or opinions that might be expressed in the written work do not necessarily represent the views of the U.S. Government. The publisher acknowledges that the U.S. Government retains a non-exclusive, paid-up, irrevocable, world-wide license to publish or reproduce the published form of this written work or allow others to do so, for U.S. Government purposes. The DOE will provide public access to results of federally sponsored research in accordance with the DOE Public Access Plan.
\end{acknowledgement}

\bibliography{refs}

\end{document}